\documentclass[pre,nopacs,twocolumn,preprintnumbers,notitlepage,amsmath,amssymb,groupedaddress]{revtex4-2}
\usepackage{amssymb}
\usepackage{amsmath}

\usepackage{color}
\usepackage[pdftex]{graphicx}

\usepackage{soul}

\usepackage{mathtools}

\usepackage{makecell}

\usepackage[dvipsnames]{xcolor}

\definecolor{darkblue}{rgb}{0,0,0.6}
\definecolor{darkred}{rgb}{0.6,0,0}
\usepackage[colorlinks=true,urlcolor=darkblue,citecolor=darkblue,linkcolor=darkred]{hyperref}

\usepackage{enumerate}
\usepackage{mathtools}
\usepackage{stmaryrd}

\usepackage{enumitem}

\usepackage{textcomp}
\usepackage{gensymb}

\newcommand{\moy}[1]{\left\langle #1 \right\rangle}

\newcommand{\ex}[1]{\mathrm{e}^{#1}}

\newcommand{\dd}[0]{\mathrm{d}}

\newcommand{\ee}[0]{\boldsymbol{e}}
\newcommand{\rr}[0]{\boldsymbol{r}}

\newcommand{\vv}[0]{\boldsymbol{v}}

\newcommand{\pp}[0]{\boldsymbol{p}}

\newcommand{\kB}[0]{k_{\mathrm{B}}}

\newcommand{\nn}[0]{{\boldsymbol{n}}}

\newcommand{\A}[0]{\text{A}}
\newcommand{\B}[0]{\text{B}}
\newcommand{\C}[0]{\text{C}}

\begin{document}

\title{Spontaneous propulsion of an isotropic colloid in a phase-separating environment}
\author{Jeanne Decayeux}
\author{Vincent Dahirel}
\author{Marie Jardat}
\author{Pierre Illien}

\affiliation{Sorbonne Universit\'e, CNRS, Laboratoire PHENIX (Physicochimie des Electrolytes et Nanosyst\`emes Interfaciaux), 4 place Jussieu, 75005 Paris, France}

\date{\today}

\begin{abstract}



The motion of active colloids is generally achieved through their anisotropy, as exemplified by Janus colloids. Recently, there was a growing interest in the propulsion of isotropic colloids, which requires some local symmetry breaking. Although several mechanisms for such propulsion were proposed, little is known about the role played by the interactions within the environment of the colloid, which can have a dramatic effect on its propulsion. Here, we propose a minimal model of an isotropic colloid in a bath of solute particles that interact with each other. These interactions lead to a spontaneous phase transition close to the colloid, to directed motion of the colloid over very long timescales and to significantly enhanced diffusion, in spite of the crowding induced by solute particles. We determine the range of parameters where this effect is observable in the model, and we propose an effective Langevin equation that accounts for it and allows one to determine the different contributions at stake in self-propulsion and enhanced diffusion.

\end{abstract}

\maketitle

\section{Introduction}

 Synthetic self-propelled particles, like active colloids, have been the subject of numerous theoretical and experimental studies during the past decades \cite{Bechinger2016, Zottl2016a, Illien2017}. Among the possible routes to locomotion, the design of \emph{anisotropic} colloids, which interact with self-generated gradients of solute concentration, temperature, or electric fields, has been particularly fruitful. This is exemplified by Janus colloids, whose hemispheres have different surface properties, for instance a catalytic and a non-catalytic one~\cite{Ebbens2010}.

There was a more recent interest in the propulsion of \emph{isotropic} colloids. It was demonstrated that built-in asymmetry of the colloid is actually not necessary to achieve directed motion over long timescales, and that a spontaneous polarization of its environment can be sufficient. For instance, one can consider isotropic colloids, which interact with solute particles that are continuously and isotropically emitted from their surface, in such a way that the number of solute particles is not conserved. Spontaneous fluctuations in the solute density field can yield transient anomalous diffusion of the colloid and enhanced diffusion \cite{Golestanian2009, Valeriani2013,Golestanian2019}. Spontaneous symmetry breaking can also arise from the nonlinear coupling between the solute density and the flows at the surface of the colloid \cite{Rednikov1994,Michelin2013,Michelin2014,Hu2019} -- an effect which was evidenced experimentally with large water droplets in an oil-surfactant medium \cite{Thutupalli2011,Izri2014,Herminghaus2014,Maass2016a,Illien2020}. Alternatively, models where the number of solute particles is conserved were considered. If the surface of the colloid plays the role of a catalyst, and if the colloid interacts differently with the reactants and the products, propulsion can also be achieved under suitable conditions \cite{DeBuyl2013a}.

\begin{figure}[b]
\begin{center}
\includegraphics[width=\columnwidth]{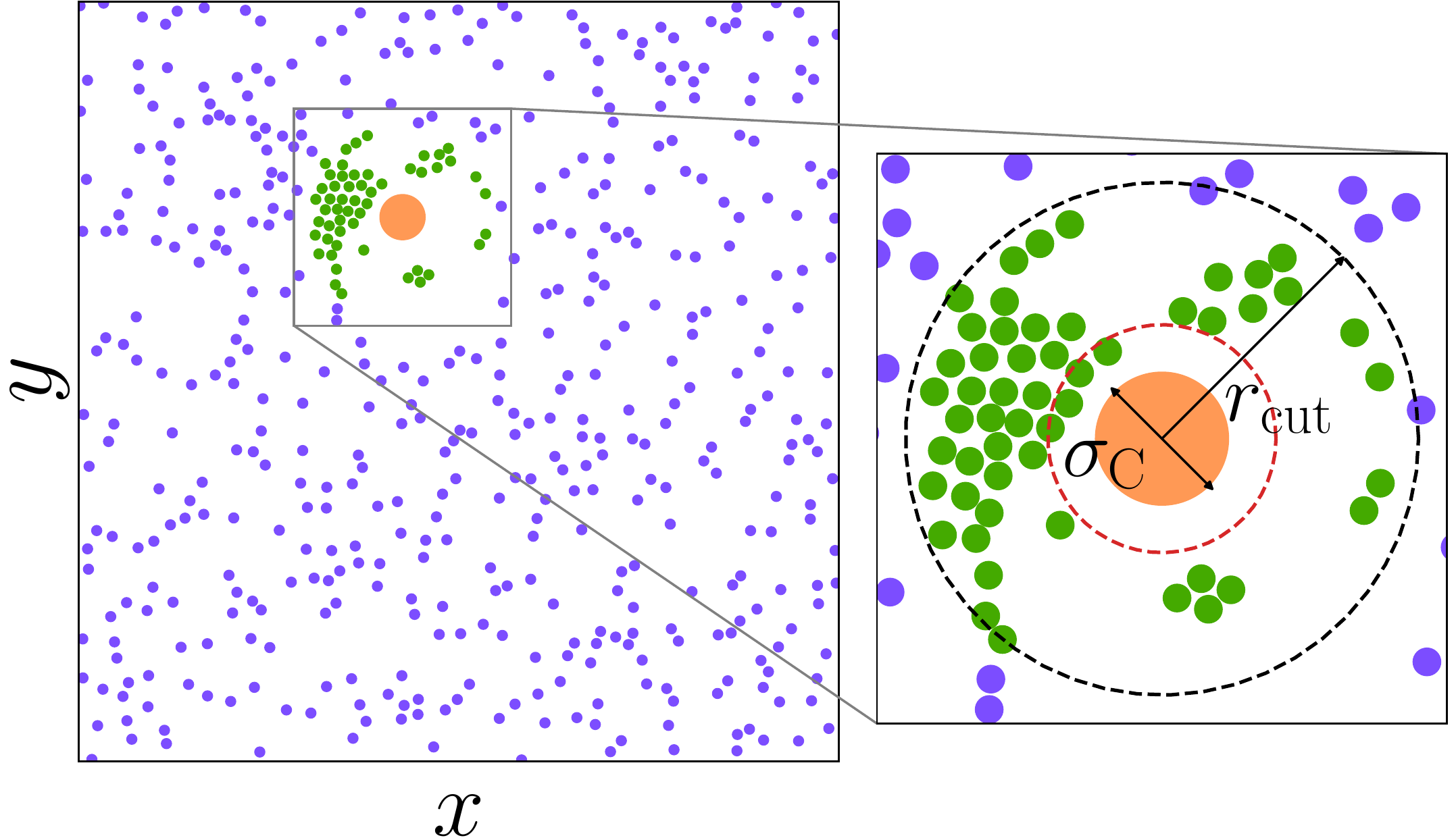}
\caption{Snapshot of the system : $N=500$ solute particles and a colloid $\mathrm{C}$ (orange) in a periodic square box. Zoom on the colloid where the reaction $\mathrm{A}+\mathrm{C} \to \mathrm{B}+\mathrm{C}$ takes place in the reaction area of radius $r_\text{cut}$. A particles are in violet, B in green. A particles interact with purely repulsive interactions, whereas the interactions between B particles are attractive (Lennard-Jones). The interactions between the colloid and the solute particles are repulsive. The red circle represents the domain $\mathcal{P}$ within which the colloid may interact directly with solute particles.}
\label{sketch}
\end{center}
\end{figure}

However, in the modeling of these isotropic self-propelled colloids, the interactions between the solute particles are generally not taken into account, although they play a significant role on the propulsion mechanism, and on the displacement of the colloid when self-propulsion occurs. This aspect becomes particularly important when such processes take place in dense environments, for instance in confined geometries or in the intracellular medium. In the later case, interactions can lead to liquid-liquid phase separations and thus to strong discontinuities in solute density gradients \cite{Hyman2014}.  However, if the solute density increases significantly in the vicinity of a colloidal particle,  solute particles may as well slow down its diffusion, as in a crowded medium. Therefore, it is necessary to adopt a finer level of description for the environment of active colloids, in order to understand the effect of solute-solute interactions on their dynamics, especially when these interactions may drive a phase transition. { Local phase separation was previously used to induce the motion of  colloids: when illuminated by light, gold-capped Janus colloids can trigger a local asymmetric demixing of a binary water-lutidine mixture, responsible for self-diffusiophoretic motion \cite{Volpe2011, Buttinoni2013}. However, our goal here is to study the self-propulsion of an \emph{isotropic} colloid, in the particular situation where the particles whose density fluctuations are responsible for its displacement, that we will call thereafter `solute particles', have a size comparable to that of the colloid and therefore induce crowding that may hinder the displacement of the colloid.}

To this end, we propose and numerically study a model for an isotropic colloid in a bath of solute particles that interact with each other. We investigate how the interactions within the environment of the colloid -- and not only those between the colloid and its environment -- control the propulsion of the colloid. Our strategy consists in designing the simplest out-of-equilibrium model where interactions can lead to a local phase transition close to the colloid.
Far from the colloid, the solute particles are of type A, and interact via purely repulsive interactions. Within a given cutoff distance from the center of the colloid, the reaction A $\to$ B takes place, and we assume that interactions between the solute particles of type B are attractive (see Fig. \ref{sketch}). If the attraction is strong enough, there can be a local phase transition of B. The two phases can coexist for a long time in the reaction area, {thus making the environment of the colloid strongly inhomogeneous}.
Such minimal model could represent a variety of biological systems, including 2-state proteins (A/B), whose conformation and/or phosphorylation state changes close to a larger microscopic structure, such as a ribosome or a vesicle.

We show that this local phase transition and the resulting {inhomogeneities} trigger self-propulsion of the colloid, and we quantify the resulting enhanced diffusion. Finally, relying on an analysis of the Brownian dynamics trajectories, we propose an effective Langevin equation to describe the dynamics of the colloid and its propulsion mechanism. This effective Langevin equation, which is derived from the microscopic dynamics, is compared to the equations of motion which are usually postulated in active matter theory.

\section{Model}

We consider a two-dimensional suspension of $N$ solute particles of diameter $\sigma_\A$ and 1 colloid of diameter $\sigma_\C$ embedded in an implicit solvent, in a square box of size $L$, with periodic boundary conditions. $\sigma_\A$ is chosen as the unit length, and the diameter of the colloid is $\sigma_\C=5\sigma_\A$. The number density of the solute is $\rho = N/L^2$, and will be fixed to $0.1$ (with $N=500$ particles) in all the simulations presented here. This corresponds to a solute surface fraction $\phi \simeq 0.079$. When a particle of type A is at a distance smaller than a cutoff distance $r_\text{cut}$ from the center of the colloid, it becomes $\B$ with rate $k_{\A\B}$ (Fig. \ref{sketch}). Conversely, outside this reactive area, $\B$ transforms back into $\A$ with rate $k_{\B\A}$. This reverse reaction maintains the system out-of-equilibrium, and mimics the flux of B away from the colloid in a system where the A species remain predominant. In all the simulations presented in the main text, we take  $k_{\A\B}=k_{\B\A}=10 \tau^{-1}$, where $\tau=\sigma_\text{A}^2/D_\text{A}$ is the typical time taken by a solute molecule to diffuse over its own size, and is chosen to be the unit time of the problem. In this way, the typical $\A\leftrightarrow \B$ conversion times are small compared to the other timescales of the problem. The C--A, C--B and A--A interactions are purely repulsive and are described with the Weeks-Chandler-Andersen (WCA) potential ${U_{\text{WCA}} (r_{ij}) }= 4\varepsilon' \left[ \left(\frac{d_{ij}}{r_{ij}}\right)^{12}-\left(\frac{d_{ij}}{r_{ij}}\right)^{6}  \right] +\varepsilon'$ where $r_{ij}$ is the distance between particles $i$ and $j$, $d_{ij}=(\sigma_\A+\sigma_C)/2$ if $i$ or $j$ is the colloid, and $d_{ij}=\sigma_\A$ otherwise \cite{Weeks1971}. The WCA potential is equal to zero for $r \geq 2^{1/6}d_{ij}$. The B particles have the same diameter as A, but interact with each other via a Lennard-Jones (LJ) potential, i.e. with an attractive part: ${U_{\text{LJ}} (r_{ij}) }= 4\varepsilon \left[ \left(\frac{\sigma_\A}{r_{ij}}\right)^{12}-\left(\frac{\sigma_\A}{r_{ij}}\right)^{6}  \right]  $. We set $\varepsilon'=10 \kB T$, and $\varepsilon$, which tunes the intensity of the attraction between B particles, varies from $\kB T$ to $3 \kB T$, which are typical values for non-specific protein-protein interactions \cite{Pellicane}. We do not aim at considering the effect of long-range interactions in this system, therefore we impose the LJ potential to vanish for $r \geq 2.5 d_{ij}$.

We simulate the system using Brownian dynamics \cite{Frenkel}. The positions of each of the $N+1$ particles in the system satisfy the overdamped Langevin equations:
\begin{equation}
\label{ }
\frac{\dd\rr_i}{\dd t} = - \frac{D_i}{\kB T} \sum_{j \neq i }\nabla U_{ij}(|\rr_i-\rr_j|) + \sqrt{2 D_i} \boldsymbol{\eta}_i(t),
\end{equation}
where $D_i \propto 1/\sigma_i$ is the bare diffusion coefficient of particle $i$, and $\boldsymbol{\eta}_i(t)$ is a white noise such that $\moy{\eta_{i,\alpha}(t)}=0$ and $\moy{\eta_{i,\alpha}(t)\eta_{j,\beta}(t')}=2D_i \delta_{ij} \delta_{\alpha\beta} \delta (t-t')$ for any components $\alpha,\beta = x$ or $y$. We integrate them with a forward Euler scheme (Appendix \ref{app:algorithm}). 
All the equations are made dimensionless accordingly. The integration timestep $\delta t$ varies between $10^{-4}\tau$ and $10^{-5}\tau$, and we integrate the equations over at least $1.5\cdot 10^6$ timesteps. 
The values of all the simulation parameters are given in Appendix \ref{app:algorithm}.

\begin{figure}
\begin{center}
\includegraphics[width=\columnwidth]{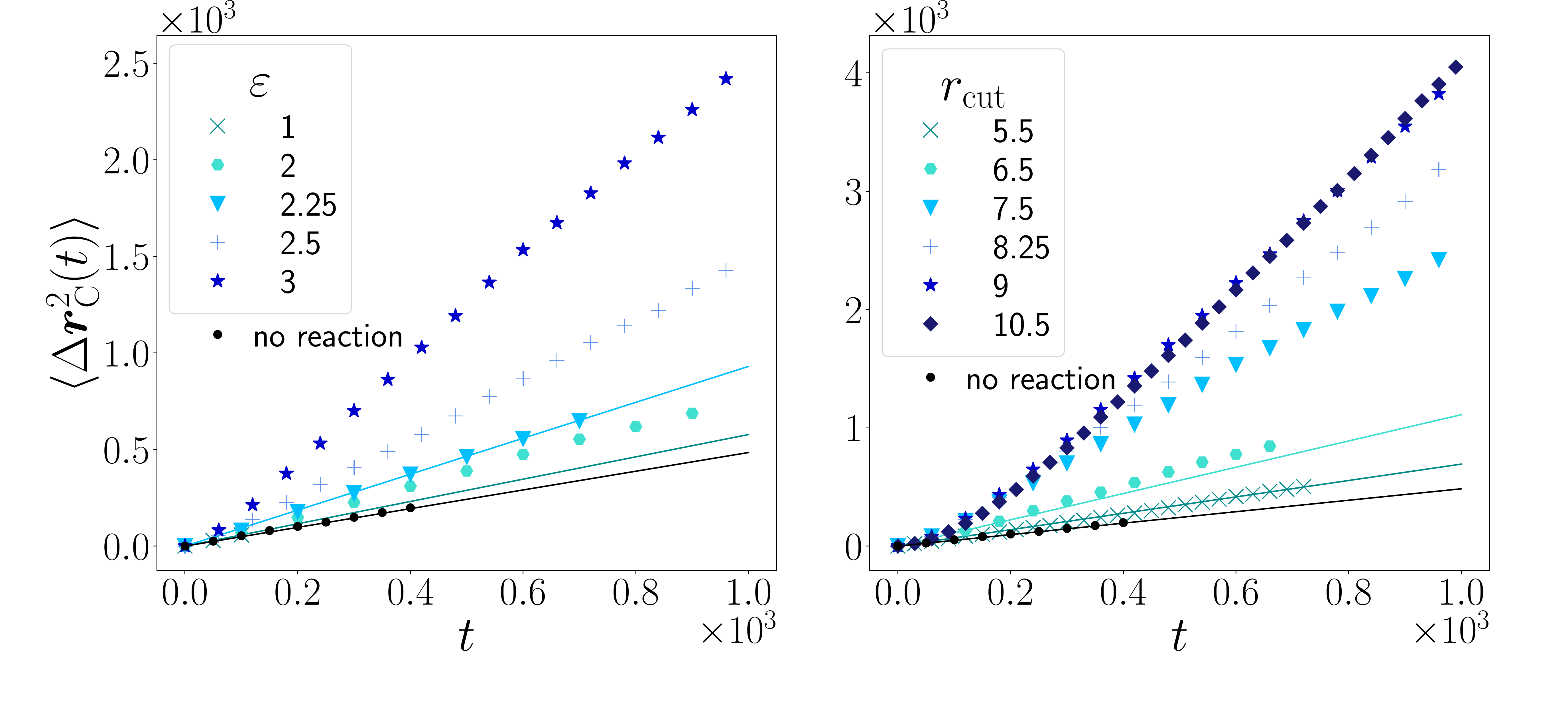}
\caption{Mean squared displacement of the colloid for fixed $r_\text{cut}=7.5$ and different values of $\varepsilon$ (left); and fixed $\varepsilon=3$ and different values of $r_\text{cut}$ (right). Solid lines are guides for the eye.}
\label{MSDfig}
\end{center}
\end{figure}

\section{Enhanced diffusion}

 We observe that, when the reaction $\A\to \B$ takes place in the vicinity of the colloid, its diffusion can be significantly enhanced compared to its equilibrium value, in spite of the crowding imposed by the solute particles. In order to quantify the diffusion enhancement, we calculate the mean squared displacement (MSD) of the colloid $\Delta \rr_\text{C} ^2(t) = \moy{[\rr_\text{C}(t)-\rr_\text{C}(0)]^2}$, where the average runs over initial conditions and noise realizations. Its dependence on time for different sizes of the reaction area is represented in Fig. \ref{MSDfig}. We observe that, at long times, the slope of the MSD increases with the size of the reaction area $r_\text{cut}$, and with the intensity of the Lennard-Jones potential $\varepsilon$. {We define the diffusion coefficient as $D\equiv \lim_{t\to\infty}\Delta \rr_\text{C} ^2(t)/4t $, and we show in Appendix \ref{app:D_rcut_eps} the dependence of $D$ over $r_\text{cut}$ and $\varepsilon$. In particular, we measure $D_{\varepsilon=3}/D_\text{eq}\simeq 5.5$ (for $r_\text{cut}=7.5$, Fig. \ref{MSDfig} left), where $D_\text{eq}$ is the reference diffusion coefficient without reaction,  and $D_{r_\text{cut}=10.5}/D_\text{eq}\simeq 9.8$ (for $\varepsilon=3$, Fig. \ref{MSDfig} right).} {Therefore, the diffusion coefficient of the colloid can be increased up to ten-fold when the colloid catalyzes the A$\to$B reaction. This diffusion enhancement, although it is smaller than those typically observed for anisotropic colloids, is particularly significant given that the colloid is isotropic, that activity is not fuelled by a significant external energy input, and that this enhancement occurs in spite of the crowding induced by solute particles, that would on the contrary tend to hinder the displacement of the colloid in a purely passive system.}

We also compare the situation where all the solute particles are of type B and interact via the LJ potential (i.e. the limit of $r_\text{cut}\to\infty$). We observe that $D_{r_\text{cut}=\infty}$ becomes comparable to $D_\text{eq}$ ($D_{r_\text{cut}=\infty}/D_\text{eq}\simeq 1.4$ for $\varepsilon=3$) {\footnote{{If $r_\text{cut}\to\infty$, all the solute particles are of type B, and a phase transition occurs in the whole bulk, and not only in the reaction area. The fluctuations in the density of B particles are not confined to a small region of space anymore, and are not localized at the close vicinity of the colloid, therefore preventing the propulsion mechanism from occurring.}}}. This suggests that, interestingly, for a fixed value of the parameter  $\varepsilon$, there exists a value of $r_\text{cut}$ that optimizes the diffusion coefficient of the colloid. {Finally, when $\varepsilon$ is too large, we expect the B solute particles to form a dense crystal around the colloid and to significantly hinder the displacement of the colloid {(Appendix~\ref{app:crystal})}.}

Qualitatively, the diffusion enhancement can be attributed to the following mechanism. When $\varepsilon$ and/or $r_\text{cut}$ are large enough, the B particles present in the reactive area around the colloid attract each other and form a cluster (Fig. \ref{sketch}), but this cluster does not fully fill the reactive area. The colloid is pushed away from the cluster due to its repulsive interactions with the solute particles. If the cluster keeps the same orientation relatively to the colloid for a sufficiently long time, this results in a propulsion of the colloid, that eventually crosses over to enhanced diffusion for observation times larger than the persistence time of the cluster orientation.

The signature of this propulsion mechanism can be seen when plotting the MSD of the colloid in a log-log scale [Fig. \ref{autocorr}(b)]. Three successive regimes can be identified. At short times, the colloid has a diffusive behaviour. At times $\gtrsim 10\tau$, the motion is almost ballistic, which is a signature of the self-propulsion of the colloid. Finally, at times $\gtrsim 100\tau$, the MSD crosses over to an ultimate diffusive regime, with a significant diffusion enhancement.

\section{Effective equation of motion}

In order to get a better insight into the self-propulsion of the colloid, we aim at coarse-graining the microscopic dynamics and writing an effective Langevin equation for the position of the colloid. We define $\pp=\sum_{i\in\mathcal{P}} [\rr_i(t) - \rr_\text{C}(t)]$, where $\mathcal{P}$ is the circular zone around the colloid where it may interact directly with solute particles (see Fig. \ref{sketch}). $\pp$  represents the polarization of solute particles around the colloid. We write the velocity of the colloid under the form 
 \begin{equation}
        \label{effective_Langevin_petitp}
        \frac{\dd}{\dd t}\rr_\text{C} = - K \pp + \boldsymbol{\xi}.
\end{equation}
The first term $-K\pp$ represents the direct interactions of the colloid with nearby solute particles, and plays the role of an effective `active force', originating from the local polarization of the environment of the colloid. The second term $\boldsymbol{\xi}$ is an effective noise term, {which is built with the constraint that its fluctuations are faster than those of the active force. The time scale characterizing the active force is defined below.} 

Integrating Eq. \eqref{effective_Langevin_petitp}, squaring it, and averaging over realizations yields the MSD of the colloid, which reads
    \begin{align}
        \moy{[\rr_\text{C}(t)-\rr_\text{C}(0)]^2} =& K^2 \Delta_{\pp\pp}(t) + \Delta_{\boldsymbol{\xi}\boldsymbol{\xi}}(t) \nonumber\\
        &-K[\Delta_{\pp\boldsymbol{\xi}}(t)+\Delta_{\boldsymbol{\xi}\pp}(t)],
        \label{eq_msd_contrib}
    \end{align}
where we define $\Delta_{\boldsymbol{a}\boldsymbol{b}}(t)\equiv \int_0^t\dd t'\int_0^t \dd t''   \moy{\boldsymbol{a}(t')\cdot\boldsymbol{b}(t'')}$. The MSD of the colloid is therefore written in terms of the integrals of different correlation functions.  

\begin{figure*}
\includegraphics[width=17.8cm]{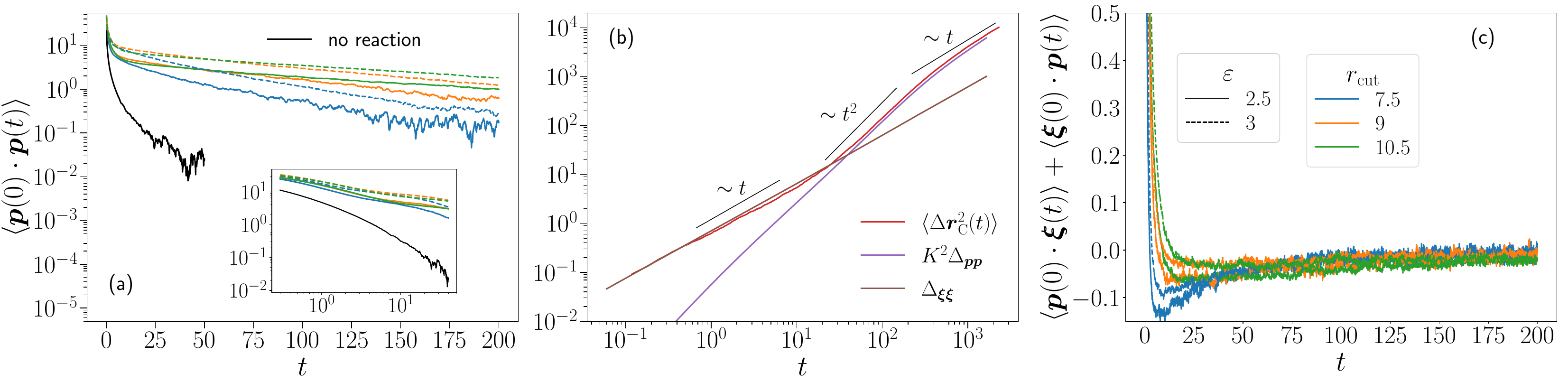}
\caption{(a) Autocorrelation of the polarization vector $\pp$, and (c) cross-correlations between $\pp$ and $\boldsymbol{\xi}$ as function of time for different values of the parameters $\varepsilon$ and $r_\text{cut}$. The legend is the same for both plots and given on panel (c). In the absence of reaction, the autocorrelation of $\pp$ decreases very fast and oscillates around zero. Inset of (a): Zoom on the short-time dynamics, represented in a log-log scale. (b) Mean squared displacement of the colloid as a function of time, and contributions defined in Eq. \eqref{eq_msd_contrib}, for $\varepsilon=3$ and $r_\text{cut}=10.5$.  The parameters for the analysis of the Brownian dynamics trajectories (b) are $K=0.0425 \tau^{-1}$ and $\tau_0 = 0.3\tau_p=49.5\tau$.}

\label{autocorr}
\end{figure*}

In order to identify the different contributions to this MSD, we first study the autocorrelation of the polarization $\langle \pp(0)\cdot \pp(t) \rangle$, which is represented on Fig. \ref{autocorr}(a) for different sets of parameters $\varepsilon$ and $r_\text{cut}$. When the reaction $\A \to \B$ takes place, this autocorrelation function typically displays two regimes: a power-law decay at short times, and exponential decay at long times [Fig. \ref{autocorr}(a)]. Interestingly, the exponential tail is not observed in the absence of reaction, i.e. when the colloid is surrounded by a homogeneous suspension of A particles, so that this exponential decay can then be seen as a signature of activity. We define a time $\tau_p$, which characterizes the persistence of the orientation of $\pp$, and which is such that  ${\langle \pp(0)\cdot\pp(t) \rangle \propto \ex{-t/\tau_p}} $ at times sufficiently large (see Appendix \ref{app:tau_p} for the values of $\tau_p$ associated to the set of parameters in Fig. \ref{autocorr}). The persistence time $\tau_p$ is an increasing function of $r_\text{cut}$ for a fixed value of $\varepsilon$. Finally, computing $\Delta_{\pp\pp}(t)\equiv \int_0^t\dd t'\int_0^t \dd t''   \moy{\pp(t')\cdot\pp(t'')}$ from the numerical data shows that the contribution from the autocorrelation of the solute polarization is responsible for the transient ballistic behavior of the colloid [Fig. \ref{autocorr}(b)].

We then study the contribution to the MSD coming from the correlations of $\boldsymbol{\xi}$, which is calculated from the trajectories as $\boldsymbol{\xi} = K \pp + \frac{\dd \rr_\text{C}}{\dd t}$.  We evaluate the derivative of the position as $\frac{\dd \rr_\text{C}}{\dd t} \simeq\frac{\Delta \rr_\text{C}}{\Delta t}$, with a sufficiently small timestep $\Delta t$. The coefficient $K$ is estimated by averaging Eq. \eqref{effective_Langevin_petitp} over a timescale $\tau_0$ which is sufficiently large to yield $\moy{\boldsymbol{\xi}}_{\tau_0}\simeq 0$ (where $\moy{\cdot}_{\tau_0}$ is a running average of duration $\tau_0$), but that remains small or comparable to the persistence time of the polarization $\tau_p$.
In this way, $\pp$ and $\vv$ remain approximately aligned over the timescale $\tau_0$. From the simulations, $K$ is measured as $K = \moy{K(t)}_t$ with $K(t)= -\moy{\pp(t)}_{\tau_0}\cdot\moy{\vv(t)}_{\tau_0}/|\moy{\pp(t)}_{\tau_0}|^2$, where $\vv = \frac{\Delta \rr_C}{\Delta t}$ is the instantaneous velocity of the colloid. 
We fix $\tau_0\simeq 0.4\tau_p$. With this choice, we observe that: (i) the contribution of the cross-correlations of $\pp$ and $\boldsymbol{\xi}$  [Fig. \ref{autocorr}(c)] to the MSD of the colloid (last term in the rhs of Eq. \eqref{effective_Langevin_petitp}) is negligible {(Appendix \ref{app:G})}; (ii) the autocorrelation of $\boldsymbol{\xi}$ only varies on a timescale very small compared to all the other timescales of the problem, and in particular to $\tau_p$ (Appendix \ref{app:autocorr_xi}). $K$ can also be estimated from analytical considerations, and we provide an order of magnitude estimate which matches quantitatively the numerical estimate (Appendix \ref{app:K}).

{Eq. \eqref{effective_Langevin_petitp} can be understood as an effective Langevin equation for the colloid, where the correlations between $\pp$ and $\boldsymbol{\xi}$ do not contribute to the MSD of the colloid.}
This equation separates the effect of the solute particles in two contributions that participate almost independently to the diffusion of the colloid: an effective active force which directly controls the motion of the colloid, and an effective bath. Finally, we can compute $\Delta_{\boldsymbol{\xi}\boldsymbol{\xi}}$ from the autocorrelation function of $\boldsymbol{\xi}$. This contribution is represented on Fig. \ref{autocorr}(b). It remains linear at all times 
and dominates the dynamics of the colloid at short times and as long as the effect of local polarization is not felt by the colloid. At times longer than $\gtrsim 10\tau$, the colloid begins to self-propel, and the contribution from $\pp$ becomes dominant. 
{The MSD that can be reconstructed from the contributions we have identified perfectly matches the MSD calculated directly from the $N$-body simulations [Fig. \ref{autocorr}(b)]. We show in Appendix \ref{app:G} that the cross-correlations between $\pp$ and $\boldsymbol{\xi}$ lead to a negligible contribution to the MSD. Remarkably, the contribution $\Delta_{\pp\pp}(t)$ perfectly matches the MSD computed from simulation at long times, and the contribution $\Delta_{\boldsymbol{\xi}\boldsymbol{\xi}}(t)$  perfectly matches the MSD computed from simulation at short times. }

As a final remark, we comment on the relation between our effective Langevin equation [Eq. \eqref{effective_Langevin_petitp}] and the usual description of active Brownian particles (ABP), which are widely used in active matter theory \cite{Bechinger2016}. The dynamics of these particles is typically described by an overdamped Langevin equation of the form $\dot{\rr}=v_0\nn+\boldsymbol{\zeta}$, where $v_0$ is the propulsion velocity, $\boldsymbol{\zeta}$ is a white noise term, and $\nn$ is the orientation of the particle, which fluctuates with time. In most cases, this orientation is assumed to have exponential correlations: $\langle\nn(0)\cdot\nn(t)\rangle\sim\ex{-D_r t}$. In this perspective, it is interesting to notice that Eq. \eqref{effective_Langevin_petitp} bears a similar structure. However, and importantly, this equation and the parameters $-K$ and $\tau_p$ (which play roles analogous to $v_0$ and $D_r^{-1}$ in ABP models), emerge from microscopic considerations and are not postulated \emph{a priori}. Moreover, our analysis also provides an example for a propulsion mechanism where the orientation dynamics  is more complicated than purely exponential, with a combination of different regimes, as shown on Fig. \ref{autocorr}(a).

\section{Conclusion}

We have presented here a minimal model for a self-propelled isotropic colloid in a bath of solute particles. The self-propulsion relies on a local phase transition of the solute particles, which attract each other when they are close to the colloid. When two solute phases coexist in the vicinity of the colloid, its local environment becomes strongly polarized, and triggers self-propulsion over long timescales, ultimately enhancing diffusion. We determine numerically the range of parameters where this effect emerges. The originality of our model relies on the fact that we account for the interactions within the environment of the colloid, a feature which is usually absent in the theoretical modeling of isotropic active colloids, and which is here responsible for the propulsion mechanism. From the analysis of the Brownian dynamics trajectories, we propose an effective Langevin equation for the dynamics of the colloid, which is compared to the usual models of active Brownian particles.

Among the different perspectives opened by the present work, it will be particularly interesting to study: (i) such colloids in inhomogeneous environments or under confinement \cite{Lippera2020b}, which will affect the polarization of the solute particles around the colloid and its propulsion; (ii) the effect of hydrodynamic interactions, which have been shown to have a dramatic effect on the collective dynamics of active colloids \cite{Zottl2014, Matas-Navarro2014, Theers2018}; (iii) situations where the phase separation is not driven by interparticle interactions but by mixing particles in contact with different thermostats \cite{Weber2016, Grosberg2015, Ilker2020, Ilker2021}.

\appendix

\section{Brownian dynamics algorithm and implementation}
\label{app:algorithm}


The overdamped Langevin equation that describes the dynamics of a solute particle $i$ interacting with particles $j$ in the presence of a solvent bath writes
\begin{equation}
\label{overdampedlangevineq}
 \vv_i(t)= - \frac{D_i}{k_B T}\sum_{ i\neq j} \nabla U(\rr_i -\rr_j) + \sqrt{2 D } \boldsymbol{\eta}_i (t)  
\end{equation}
where $\vv_i$ is the velocity of $i$, $U$ is the pair interaction potential between solutes, $D_i$ is the diffusion coefficient of $i$ at infinite dilution and $\boldsymbol{\eta}_i$ is a white noise.
~\eqref{overdampedlangevineq} can be directly integrated with the Euler scheme into
\begin{align}
\label{euler1}
\rr_i(t+\Delta t) = &\rr_i(t) - \frac{D_i}{k_B T} \sum_{ i\neq j} \nabla U(\rr_i  -\rr_j) \Delta t  \nonumber\\
&+ \int_t^{t+\Delta t} \sqrt{2 D } \boldsymbol{\eta}_i(t') \, \mathrm dt' 
\end{align}
where $B(\Delta t)=\int_t^{t+\Delta t} \sqrt{2 D }  \boldsymbol{\eta}_i(t') \, \mathrm dt'$ is the Gaussian random variable with variance $\langle B^2 \rangle = 2 D \int_t^{t+\Delta t}  \int_t^{t+\Delta t} \langle \boldsymbol{\eta}_i(t') \boldsymbol{\eta}_i(t'') \rangle \, \mathrm dt' \, \mathrm dt''= 2 D \Delta t$.
The following equation of motion is thus iteratively used to compute the successive positions of solutes included in the square simulation box with periodic boundary conditions, starting from a random initial configuration of solute particles 
\begin{equation}
\label{euler}
\rr_i(t+\Delta t) = \rr_i(t) - \frac{D_i}{k_B T} \sum_{ i\neq j} \nabla U(\rr_i  -\rr_j) \Delta t + \sqrt{2 D \Delta t} \boldsymbol{\eta}_i
\end{equation}
As interactions are short-ranged, we use 
a cell list algorithm to compute them, that reduces the algorithm order to $N$, as described in \cite{Frenkel}.

The system consists of a large colloid $\C$ of diameter $\sigma_{\C}$ surrounded by $N$ small solutes $\A$ and $\B$ of same diameter $\sigma_{\A}$. 
$N=N_A+N_B$ is fixed to $N=500$. We take $\sigma_{\C}=5\sigma_{\A}$ and $D_{\C}=\frac{1}{5}D_{\A}$. The simulation box has the fixed size $L_{\rm box}=70\sigma_{\A}$. The colloid is assumed to catalyze isotropically the reaction A+C $\to$ B+C inside an area of radius $r_{\rm cut}$.  At each timestep of the simulation, the distance $r$ from the center of the colloid of each A particle is computed. If it is smaller than $r_{\rm cut}$,  A becomes B with a probability equal to $\Delta t\cdot k_{\A\B}$ with $\Delta t$ the simulation timestep. Also, the distances of B particles to the center of the colloid are computed at each timestep. B becomes an A particle with a probability $\Delta t\cdot k_{\B\A}$ if the distance is larger than $r_{\rm cut}$. We take $k_{\A\B}=k_{\B\A}=10\tau^{-1}$.

The simulation procedure is the following. First, the system that contains only A particles without any reaction is equilibrated for about $10 \tau$. Then, $1000$ independent configurations of this system are taken as initial configurations for runs where the chemical reaction occurs. It takes some time for the system with reaction to reach a stationary state, depending on the value of the parameters $r_{\rm cut}$ and $\varepsilon$. A stationary state is assumed to be reached when the number of B particles at a distance from the center of C smaller than $r_{\rm cut}$ is almost constant. The characteristic times $\tau_{\rm stationary}$ needed to reach that state, the values of the average number of particles in the reaction area and the total simulation times are collected in Table \ref{params} for systems where an activity was observed.

Once a stationary state is reached, the radial distribution functions, the polarization vector $\pp$ as a function of time, and the mean squared displacements of C are computed as averages over the independent realizations. The total simulation time $\tau_{\rm total}$ also depends on the parameters: it is long enough to ensure that a regular diffusion behavior is recovered in the case where activity is observed. 

\begin{table}
    \begin{tabular}{|c|c|c|c|c|c|c|}
  \hline
   \multicolumn{7}{|c|} {$\varepsilon=2$} \\ \hline
  $r_{\rm cut}$ &  5.5 & 6.5 & 7.5 & 8.25 & 9.0 & 10.5 \\
   $\tau_{\rm stationary}$ &  110 & 112 & 140 & 173 & 188 & 245 \\
 $\langle N \rangle$ &  8 & 14 & 23 & 31 & 39 & 59 \\
  $\tau_{\rm total}$ &  750 & 750 & 1350 & 1350 & 2400 & 1650 \\
    \hline
    \end{tabular}
    \begin{tabular}{|c|c|c|c|c|c|}
  \hline
   \multicolumn{6}{|c|} {$\varepsilon=2.5$} \\ \hline
    5.5 & 6.5 & 7.5 & 8.25 & 9.0 & 10.5 \\
     60 & 123 & 115 & 138 & 147 & 156 \\
  8 & 16 & 28 & 38 & 49 & 73 \\
    750 & 750 & 1500 & 1200 & 2400 & 2550 \\
    \hline
    \end{tabular}
    \begin{tabular}{|c|c|c|c|c|c|}
  \hline
   \multicolumn{6}{|c|} {$\varepsilon=3$} \\ \hline
   5.5 & 6.5 & 7.5 & 8.25 & 9.0 & 10.5 \\
     67 & 98 & 91 & 94 & 94 & 109 \\
  9 & 18 & 30 & 41 & 52 & 78 \\
   750 & 750 & 1500 & 1350 & 2400 & 2700 \\
    \hline
    \end{tabular}
    \caption{Parameters of the simulated systems in the cases where activity was observed}
    \label{params}
\end{table}



\section{Dependence of $D$ over $\varepsilon$ and $r_\text{cut}$}
\label{app:D_rcut_eps}

\begin{figure}
\includegraphics[width=0.48\columnwidth]{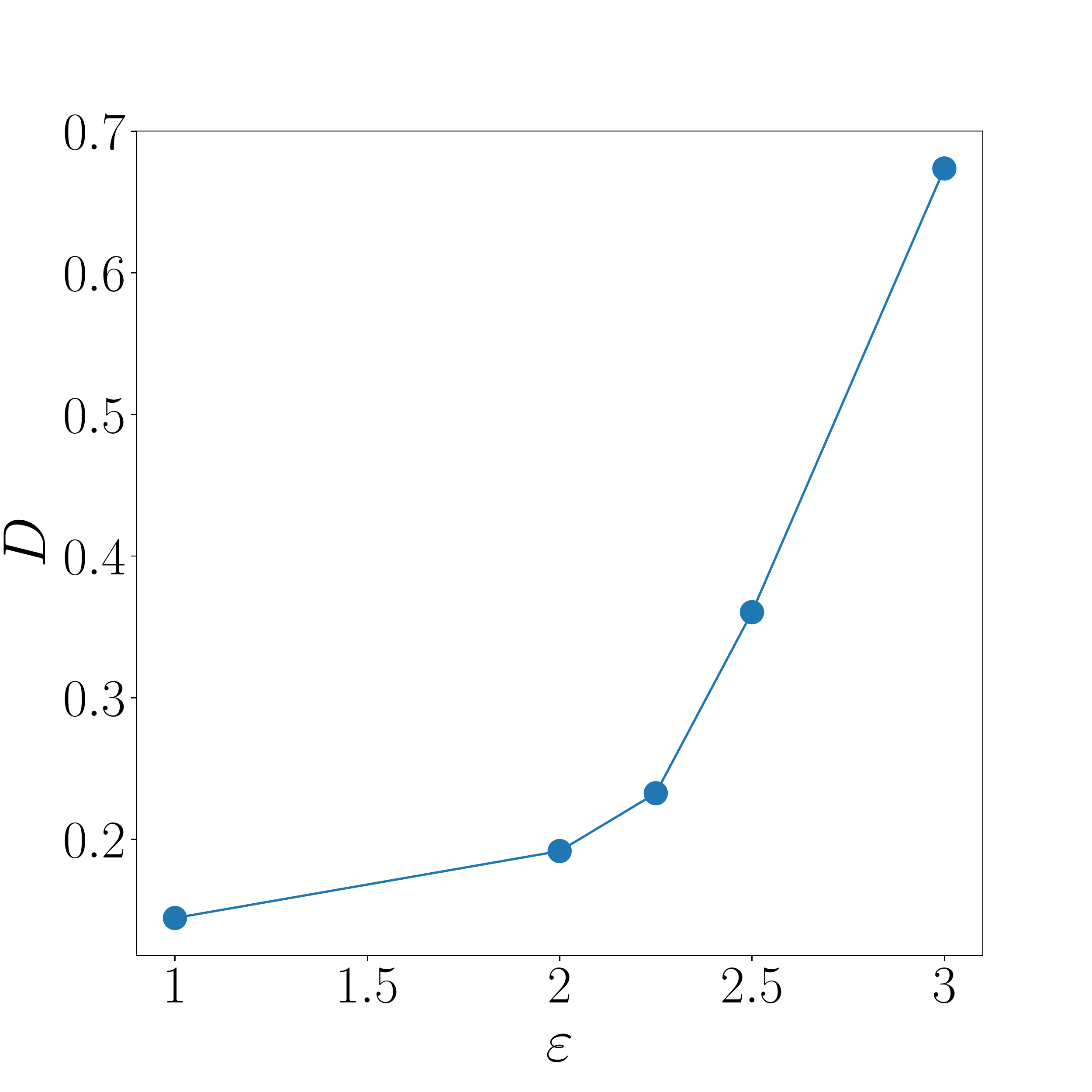}
\includegraphics[width=0.48\columnwidth]{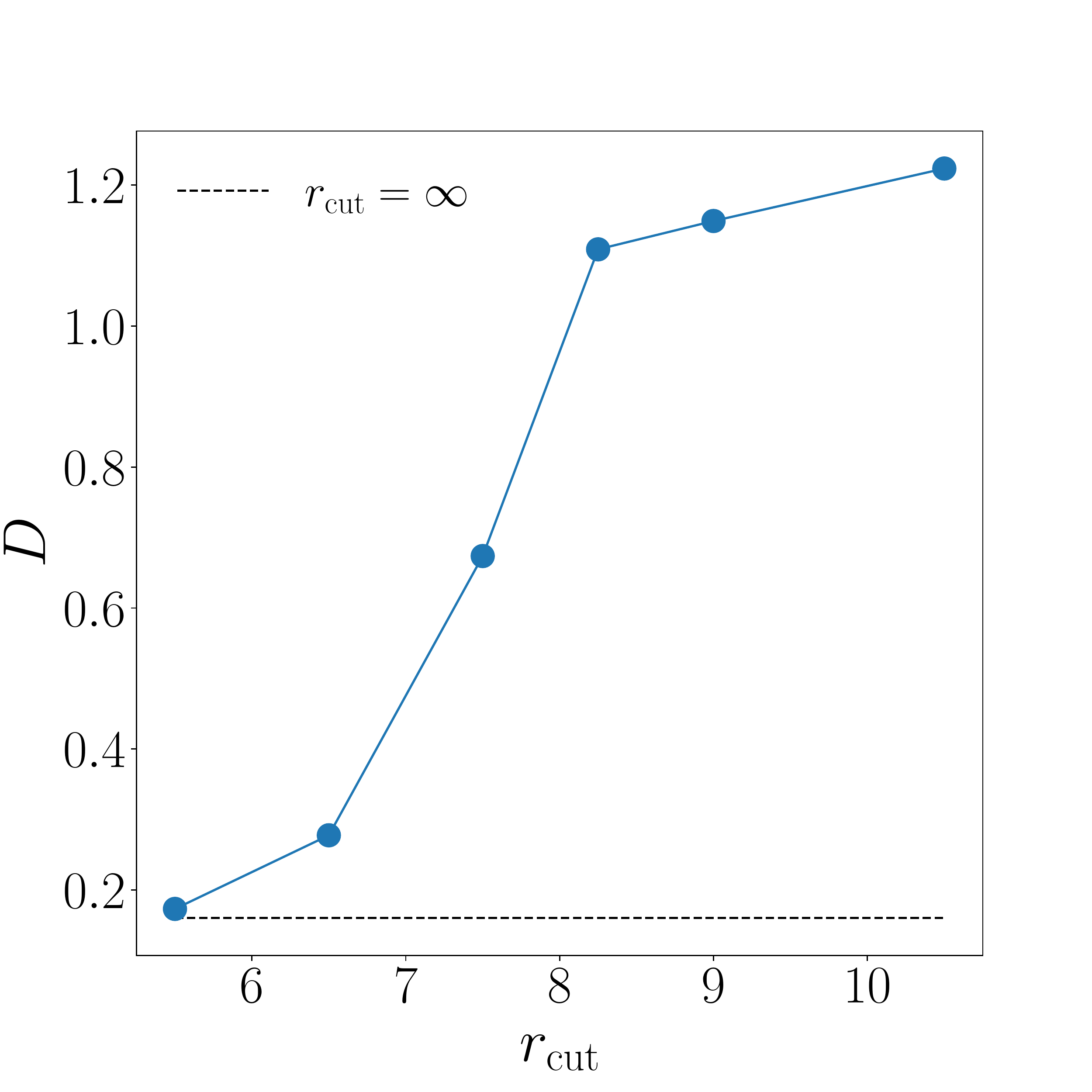}
\caption{Diffusion coefficient as a function of the parameters $\varepsilon$ and $r_\text{cut}$ (obtained from the MSD data shown on Fig. \ref{MSDfig}).}
\label{fig:D_rcut_eps}
\end{figure}

We show on Fig. \ref{fig:D_rcut_eps} the dependence of $D$ over $\varepsilon$ and $r_\text{cut}$ (the values are obtained from the MSD data shown on Fig. \ref{MSDfig}).

\section{Additional snapshot: Colloid surrounded by a dense crystal of solute}
\label{app:crystal}

\begin{figure}[!h]
\includegraphics[width=7cm]{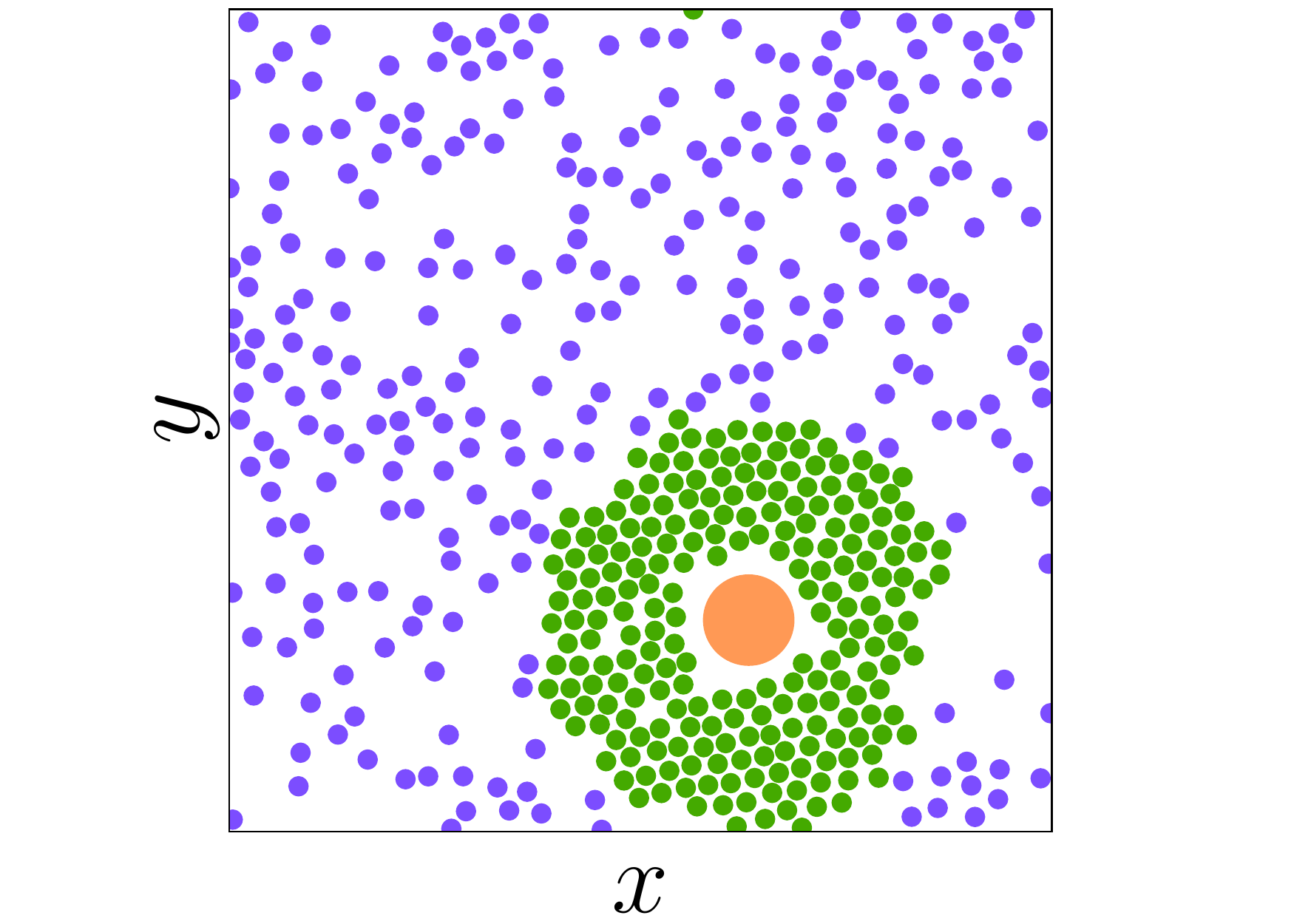}
\caption{Snapshot of a system displaying a dense cluster of solute particles around the colloid in the stationary state  ($\rho=0.3$, $\varepsilon=3$).}
\label{fig:snapshot_cristal}
\end{figure}

We expect that, for a given value of the density, when $\varepsilon$ is too large, the diffusion enhancement effect cannot be observed as the B solute particles form a dense crystal around the colloid and hinder the colloid. We show such a situation on Fig. \ref{fig:snapshot_cristal}.


\section{Values of the persistence time $\tau_p$}
\label{app:tau_p}

The values of the persistence time $\tau_p$ for the different sets of parameters used in Fig. 3(b) are given in Table \ref{table_taup}.

\begin{table}[!h]
    \begin{tabular}{|c|c|c|c|c|c|c|}
  \hline
   \multicolumn{4}{|c|} {$\varepsilon=2.5$} \\ \hline
  $r_{\rm cut}$ &  5.5 & 9 & 10.5  \\
   $\tau_p$ &  39 & 98 & 187 \\
    \hline
    \end{tabular}
        \begin{tabular}{|c|c|c|c|c|c|}
  \hline
   \multicolumn{3}{|c|} {$\varepsilon=3$} \\ \hline
     5.5 & 9 & 10.5  \\
     49 & 105 & 165 \\
    \hline
    \end{tabular}
    \caption{Values of the persistence time $\tau_p$ for the different sets of parameters used in Fig. 3(b).}
    \label{table_taup}
\end{table}


 \section{{Contributions to the MSD deduced from the effective Langevin equation}}
\label{app:G}

\begin{figure}[!h]
\includegraphics[width=9cm]{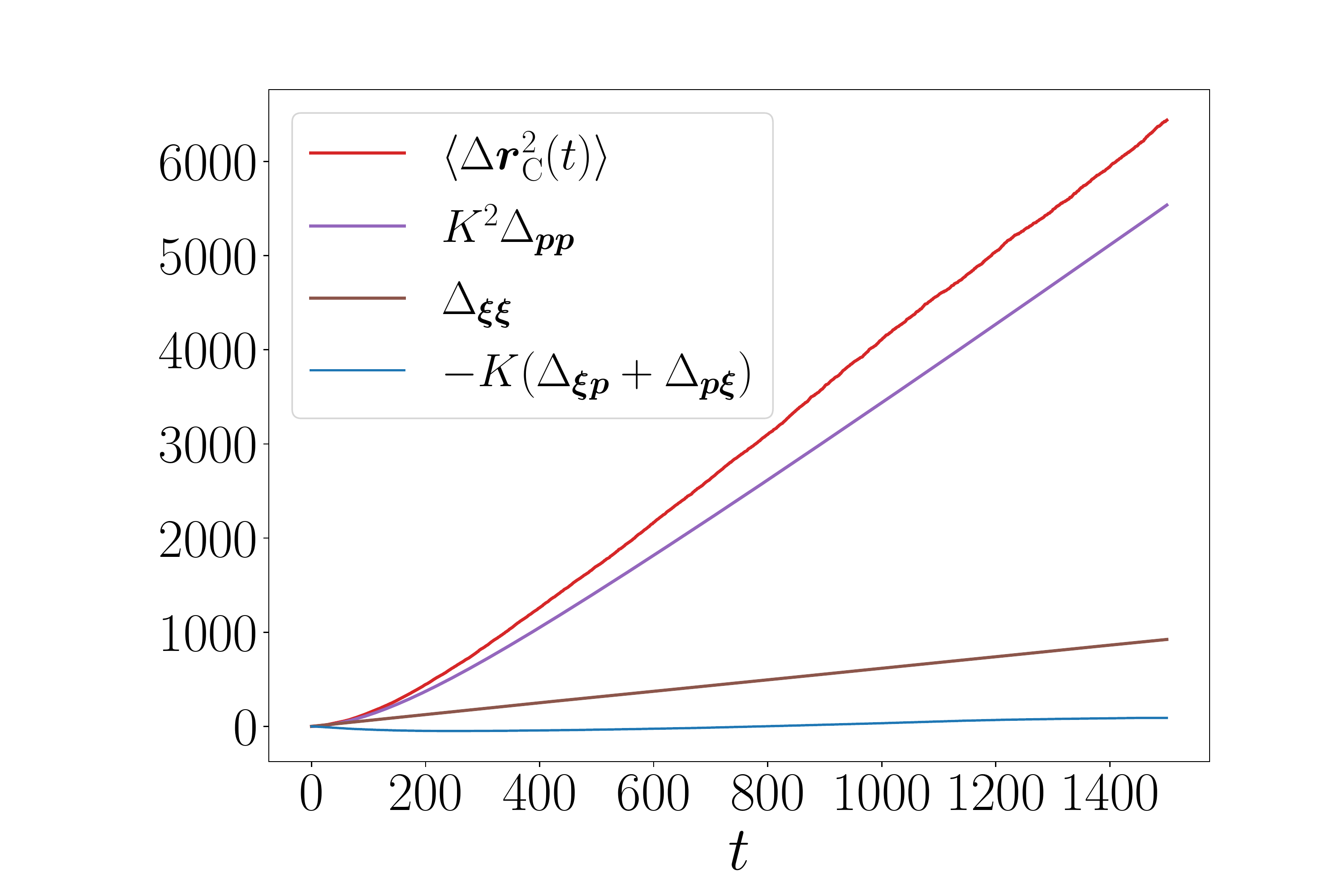}
\caption{{Mean squared displacement of the colloid as a function of time, and contributions in Eq. \eqref{eq_msd_contrib}, for $\varepsilon=3$ and $r_\text{cut}=10.5$.}}
\label{fig:lin_lin_contribution}
\end{figure}

We show on Fig. \ref{fig:lin_lin_contribution} the mean squared displacement of the colloid as a function of time, and contributions in Eq. \eqref{eq_msd_contrib}, for $\varepsilon=3$ and $r_\text{cut}=10.5$.

\section{Autocorrelation of $\boldsymbol{\xi}$}
\label{app:autocorr_xi}

\begin{figure}
\includegraphics[width=9cm]{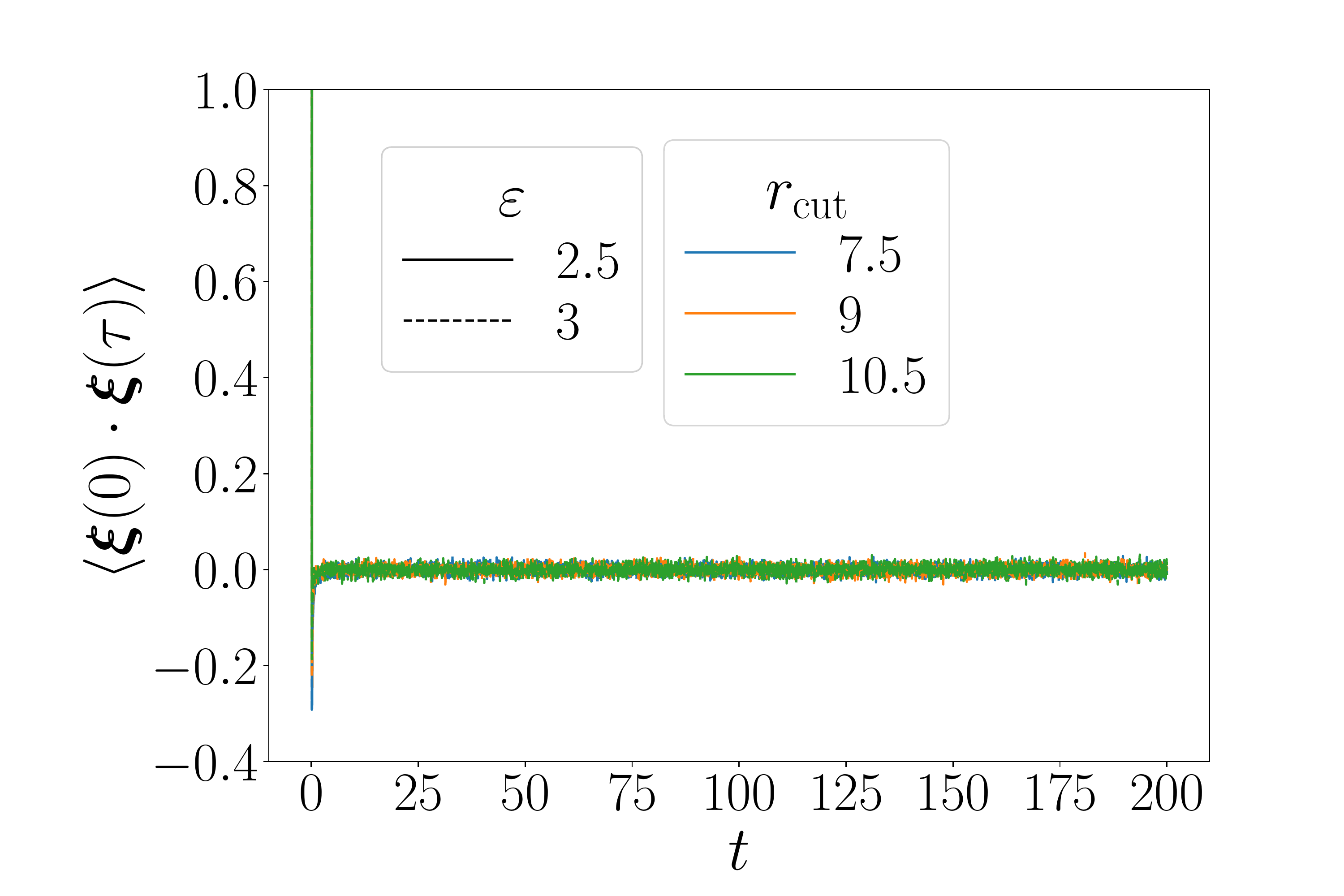}
\caption{Autocorrelation of $\boldsymbol{\xi}$ as a function of time for different values of the parameters $\varepsilon$ and $r_\text{cut}$.}
\label{autocorrelation_xi}
\end{figure}

We show on Fig. \ref{autocorrelation_xi} the autocorrelation of $\boldsymbol{\xi}$ as a function of time for different values of the parameters $\varepsilon$ and $r_\text{cut}$.

\section{Analytical estimate of the coefficient $K$}
\label{app:K}

The coefficient $K$, that appears in the effective Langevin equation in the main text [Eq. (2)] can also be estimated from analytical arguments. The velocity of the colloid (averaged over a duration comparable to the persistence time of the trajectory) can be estimated as 
\begin{equation}
\vv = \mu_\text{C} \int_{\rr \in \mathcal{P}} \dd\rr \; c(\rr)\nabla U(\rr),
\end{equation}
    where $\mu_\text{C} = {D_\text{C}}/{\kB T}$ is the mobility of the colloid, $c(\rr)$ is the concentration of solute particles measured from the centre of the colloid, and $U$ is the WCA potential acting between the colloid and the solute particles. The integral is therefore an estimate of the net force acting on the colloid. Using polar coordinates centered on the colloid, assuming that the solute concentration can be written as $c(r,\theta)\simeq \mathcal{C}(\theta)\ex{-U(r)/\kB T}$, and performing integration by parts, we find that 
    \begin{equation}
    \vv=-D_0 \left[\int_0^\infty \dd r\; (1-\ex{-U(r)/\kB T})\right] \int_{-\pi}^\pi \dd\theta\, \mathcal{C}(\theta)\ee_r.
    \end{equation}
      Similarly, the polarization can be estimated as $\pp \simeq \int_{\mathcal{P}} \dd \rr\, c(\rr)$. Under the same assumption, one gets
    \begin{equation}
    \pp \simeq \left[\int_{R}^{R+\delta}\dd r\, r^2 \ex{-U(r)/\kB T} \right] \int_{-\pi}^\pi \dd\theta\,\mathcal{C}(\theta)\ee_r.
    \end{equation}
     The coefficient $K$ can be estimated as 
    \begin{equation}
    K=\frac{D_0 \int_0^\infty \dd r\; (1-\ex{-U(r)/\kB T})}{\int_{R}^{R+\delta}\dd r\, r^2 \ex{-U(r)/\kB T} }
     \end{equation}
With the parameters used in numerical simulations, we find $K\simeq 0.037$, whose order of magnitude matches correctly the numerical observations ($K=0.0425\tau^{-1}$, see caption of Fig. 3 in the main text).


\begin{thebibliography}{32}%
\makeatletter
\providecommand \@ifxundefined [1]{%
 \@ifx{#1\undefined}
}%
\providecommand \@ifnum [1]{%
 \ifnum #1\expandafter \@firstoftwo
 \else \expandafter \@secondoftwo
 \fi
}%
\providecommand \@ifx [1]{%
 \ifx #1\expandafter \@firstoftwo
 \else \expandafter \@secondoftwo
 \fi
}%
\providecommand \natexlab [1]{#1}%
\providecommand \enquote  [1]{``#1''}%
\providecommand \bibnamefont  [1]{#1}%
\providecommand \bibfnamefont [1]{#1}%
\providecommand \citenamefont [1]{#1}%
\providecommand \href@noop [0]{\@secondoftwo}%
\providecommand \href [0]{\begingroup \@sanitize@url \@href}%
\providecommand \@href[1]{\@@startlink{#1}\@@href}%
\providecommand \@@href[1]{\endgroup#1\@@endlink}%
\providecommand \@sanitize@url [0]{\catcode `\\12\catcode `\$12\catcode
  `\&12\catcode `\#12\catcode `\^12\catcode `\_12\catcode `\%12\relax}%
\providecommand \@@startlink[1]{}%
\providecommand \@@endlink[0]{}%
\providecommand \url  [0]{\begingroup\@sanitize@url \@url }%
\providecommand \@url [1]{\endgroup\@href {#1}{\urlprefix }}%
\providecommand \urlprefix  [0]{URL }%
\providecommand \Eprint [0]{\href }%
\providecommand \doibase [0]{http://dx.doi.org/}%
\providecommand \selectlanguage [0]{\@gobble}%
\providecommand \bibinfo  [0]{\@secondoftwo}%
\providecommand \bibfield  [0]{\@secondoftwo}%
\providecommand \translation [1]{[#1]}%
\providecommand \BibitemOpen [0]{}%
\providecommand \bibitemStop [0]{}%
\providecommand \bibitemNoStop [0]{.\EOS\space}%
\providecommand \EOS [0]{\spacefactor3000\relax}%
\providecommand \BibitemShut  [1]{\csname bibitem#1\endcsname}%
\let\auto@bib@innerbib\@empty
\bibitem [{\citenamefont {Bechinger}\ \emph {et~al.}(2016)\citenamefont
  {Bechinger}, \citenamefont {{Di Leonardo}}, \citenamefont {L{\"{o}}wen},
  \citenamefont {Reichhardt}, \citenamefont {Volpe},\ and\ \citenamefont
  {Volpe}}]{Bechinger2016}%
  \BibitemOpen
  \bibfield  {author} {\bibinfo {author} {\bibfnamefont {C.}~\bibnamefont
  {Bechinger}}, \bibinfo {author} {\bibfnamefont {R.}~\bibnamefont {{Di
  Leonardo}}}, \bibinfo {author} {\bibfnamefont {H.}~\bibnamefont
  {L{\"{o}}wen}}, \bibinfo {author} {\bibfnamefont {C.}~\bibnamefont
  {Reichhardt}}, \bibinfo {author} {\bibfnamefont {G.}~\bibnamefont {Volpe}}, \
  and\ \bibinfo {author} {\bibfnamefont {G.}~\bibnamefont {Volpe}},\ }\href
  {\doibase 10.1103/RevModPhys.88.045006} {\bibfield  {journal} {\bibinfo
  {journal} {Rev. Mod. Phys.}\ }\textbf {\bibinfo {volume} {88}},\ \bibinfo
  {pages} {045006} (\bibinfo {year} {2016})}\BibitemShut {NoStop}%
\bibitem [{\citenamefont {Z{\"{o}}ttl}\ and\ \citenamefont
  {Stark}(2016)}]{Zottl2016a}%
  \BibitemOpen
  \bibfield  {author} {\bibinfo {author} {\bibfnamefont {A.}~\bibnamefont
  {Z{\"{o}}ttl}}\ and\ \bibinfo {author} {\bibfnamefont {H.}~\bibnamefont
  {Stark}},\ }\href {\doibase 10.1088/0953-8984/28/25/253001} {\bibfield
  {journal} {\bibinfo  {journal} {J. Phys. Condens. Matter}\ }\textbf {\bibinfo
  {volume} {28}},\ \bibinfo {pages} {253001} (\bibinfo {year}
  {2016})}\BibitemShut {NoStop}%
\bibitem [{\citenamefont {Illien}\ \emph {et~al.}(2017)\citenamefont {Illien},
  \citenamefont {Golestanian},\ and\ \citenamefont {Sen}}]{Illien2017}%
  \BibitemOpen
  \bibfield  {author} {\bibinfo {author} {\bibfnamefont {P.}~\bibnamefont
  {Illien}}, \bibinfo {author} {\bibfnamefont {R.}~\bibnamefont {Golestanian}},
  \ and\ \bibinfo {author} {\bibfnamefont {A.}~\bibnamefont {Sen}},\ }\href
  {\doibase 10.1039/C7CS00087A} {\bibfield  {journal} {\bibinfo  {journal}
  {Chem. Soc. Rev.}\ }\textbf {\bibinfo {volume} {46}},\ \bibinfo {pages}
  {5508} (\bibinfo {year} {2017})}\BibitemShut {NoStop}%
\bibitem [{\citenamefont {Ebbens}\ and\ \citenamefont
  {Howse}(2010)}]{Ebbens2010}%
  \BibitemOpen
  \bibfield  {author} {\bibinfo {author} {\bibfnamefont {S.~J.}\ \bibnamefont
  {Ebbens}}\ and\ \bibinfo {author} {\bibfnamefont {J.~R.}\ \bibnamefont
  {Howse}},\ }\href {\doibase 10.1039/b918598d} {\bibfield  {journal} {\bibinfo
   {journal} {Soft Matter}\ }\textbf {\bibinfo {volume} {6}},\ \bibinfo {pages}
  {726} (\bibinfo {year} {2010})}\BibitemShut {NoStop}%
\bibitem [{\citenamefont {Golestanian}(2009)}]{Golestanian2009}%
  \BibitemOpen
  \bibfield  {author} {\bibinfo {author} {\bibfnamefont {R.}~\bibnamefont
  {Golestanian}},\ }\href {\doibase 10.1103/PhysRevLett.102.188305} {\bibfield
  {journal} {\bibinfo  {journal} {Phys. Rev. Lett.}\ }\textbf {\bibinfo
  {volume} {102}},\ \bibinfo {pages} {188305} (\bibinfo {year} {2009})},\
  \Eprint {http://arxiv.org/abs/0904.3044} {arXiv:0904.3044} \BibitemShut
  {NoStop}%
\bibitem [{\citenamefont {Valeriani}\ \emph {et~al.}(2013)\citenamefont
  {Valeriani}, \citenamefont {Allen},\ and\ \citenamefont
  {Marenduzzo}}]{Valeriani2013}%
  \BibitemOpen
  \bibfield  {author} {\bibinfo {author} {\bibfnamefont {C.}~\bibnamefont
  {Valeriani}}, \bibinfo {author} {\bibfnamefont {R.~J.}\ \bibnamefont
  {Allen}}, \ and\ \bibinfo {author} {\bibfnamefont {D.}~\bibnamefont
  {Marenduzzo}},\ }\href {\doibase 10.1063/1.3428663} {\bibfield  {journal}
  {\bibinfo  {journal} {J. Chem. Phys.}\ }\textbf {\bibinfo {volume} {132}},\
  \bibinfo {pages} {204904} (\bibinfo {year} {2013})}\BibitemShut {NoStop}%
\bibitem [{\citenamefont {Golestanian}(2019)}]{Golestanian2019}%
  \BibitemOpen
  \bibfield  {author} {\bibinfo {author} {\bibfnamefont {R.}~\bibnamefont
  {Golestanian}},\ }\href {http://arxiv.org/abs/1909.03747} {\  (\bibinfo
  {year} {2019})},\ \Eprint {http://arxiv.org/abs/1909.03747}
  {arXiv:1909.03747} \BibitemShut {NoStop}%
\bibitem [{\citenamefont {Rednikov}\ \emph {et~al.}(1994)\citenamefont
  {Rednikov}, \citenamefont {Ryazantsev},\ and\ \citenamefont
  {Velarde}}]{Rednikov1994}%
  \BibitemOpen
  \bibfield  {author} {\bibinfo {author} {\bibfnamefont {A.~Y.}\ \bibnamefont
  {Rednikov}}, \bibinfo {author} {\bibfnamefont {Y.~S.}\ \bibnamefont
  {Ryazantsev}}, \ and\ \bibinfo {author} {\bibfnamefont {M.~G.}\ \bibnamefont
  {Velarde}},\ }\href {\doibase 10.1016/0017-9310(94)90036-1} {\bibfield
  {journal} {\bibinfo  {journal} {Phys. Fluids}\ }\textbf {\bibinfo {volume}
  {6}},\ \bibinfo {pages} {451} (\bibinfo {year} {1994})}\BibitemShut {NoStop}%
\bibitem [{\citenamefont {Michelin}\ \emph {et~al.}(2013)\citenamefont
  {Michelin}, \citenamefont {Lauga},\ and\ \citenamefont
  {Bartolo}}]{Michelin2013}%
  \BibitemOpen
  \bibfield  {author} {\bibinfo {author} {\bibfnamefont {S.}~\bibnamefont
  {Michelin}}, \bibinfo {author} {\bibfnamefont {E.}~\bibnamefont {Lauga}}, \
  and\ \bibinfo {author} {\bibfnamefont {D.}~\bibnamefont {Bartolo}},\
  }\href@noop {} {\bibfield  {journal} {\bibinfo  {journal} {Phys. Fluids}\
  }\textbf {\bibinfo {volume} {25}},\ \bibinfo {pages} {061701} (\bibinfo
  {year} {2013})}\BibitemShut {NoStop}%
\bibitem [{\citenamefont {Michelin}\ and\ \citenamefont
  {Lauga}(2014)}]{Michelin2014}%
  \BibitemOpen
  \bibfield  {author} {\bibinfo {author} {\bibfnamefont {S.}~\bibnamefont
  {Michelin}}\ and\ \bibinfo {author} {\bibfnamefont {E.}~\bibnamefont
  {Lauga}},\ }\href {\doibase 10.1017/jfm.2014.158} {\bibfield  {journal}
  {\bibinfo  {journal} {J. Fluid Mech}\ }\textbf {\bibinfo {volume} {747}},\
  \bibinfo {pages} {572} (\bibinfo {year} {2014})}\BibitemShut {NoStop}%
\bibitem [{\citenamefont {Hu}\ \emph {et~al.}(2019)\citenamefont {Hu},
  \citenamefont {Lin}, \citenamefont {Rafai},\ and\ \citenamefont
  {Misbah}}]{Hu2019}%
  \BibitemOpen
  \bibfield  {author} {\bibinfo {author} {\bibfnamefont {W.~F.}\ \bibnamefont
  {Hu}}, \bibinfo {author} {\bibfnamefont {T.~S.}\ \bibnamefont {Lin}},
  \bibinfo {author} {\bibfnamefont {S.}~\bibnamefont {Rafai}}, \ and\ \bibinfo
  {author} {\bibfnamefont {C.}~\bibnamefont {Misbah}},\ }\href {\doibase
  10.1103/PhysRevLett.123.238004} {\bibfield  {journal} {\bibinfo  {journal}
  {Physical Review Letters}\ }\textbf {\bibinfo {volume} {123}},\ \bibinfo
  {pages} {238004} (\bibinfo {year} {2019})}\BibitemShut {NoStop}%
\bibitem [{\citenamefont {Thutupalli}\ \emph {et~al.}(2011)\citenamefont
  {Thutupalli}, \citenamefont {Seemann},\ and\ \citenamefont
  {Herminghaus}}]{Thutupalli2011}%
  \BibitemOpen
  \bibfield  {author} {\bibinfo {author} {\bibfnamefont {S.}~\bibnamefont
  {Thutupalli}}, \bibinfo {author} {\bibfnamefont {R.}~\bibnamefont {Seemann}},
  \ and\ \bibinfo {author} {\bibfnamefont {S.}~\bibnamefont {Herminghaus}},\
  }\href {\doibase 10.1088/1367-2630/13/7/073021} {\bibfield  {journal}
  {\bibinfo  {journal} {New Journal of Physics}\ }\textbf {\bibinfo {volume}
  {13}},\ \bibinfo {pages} {073021} (\bibinfo {year} {2011})}\BibitemShut
  {NoStop}%
\bibitem [{\citenamefont {Izri}\ \emph {et~al.}(2014)\citenamefont {Izri},
  \citenamefont {{Van Der Linden}}, \citenamefont {Michelin},\ and\
  \citenamefont {Dauchot}}]{Izri2014}%
  \BibitemOpen
  \bibfield  {author} {\bibinfo {author} {\bibfnamefont {Z.}~\bibnamefont
  {Izri}}, \bibinfo {author} {\bibfnamefont {M.~N.}\ \bibnamefont {{Van Der
  Linden}}}, \bibinfo {author} {\bibfnamefont {S.}~\bibnamefont {Michelin}}, \
  and\ \bibinfo {author} {\bibfnamefont {O.}~\bibnamefont {Dauchot}},\ }\href
  {\doibase 10.1103/PhysRevLett.113.248302} {\bibfield  {journal} {\bibinfo
  {journal} {Phys. Rev. Lett.}\ }\textbf {\bibinfo {volume} {113}},\ \bibinfo
  {pages} {248302} (\bibinfo {year} {2014})}\BibitemShut {NoStop}%
\bibitem [{\citenamefont {Herminghaus}\ \emph {et~al.}(2014)\citenamefont
  {Herminghaus}, \citenamefont {Maass}, \citenamefont {Kr{\"{u}}ger},
  \citenamefont {Thutupalli}, \citenamefont {Goehring},\ and\ \citenamefont
  {Bahr}}]{Herminghaus2014}%
  \BibitemOpen
  \bibfield  {author} {\bibinfo {author} {\bibfnamefont {S.}~\bibnamefont
  {Herminghaus}}, \bibinfo {author} {\bibfnamefont {C.~C.}\ \bibnamefont
  {Maass}}, \bibinfo {author} {\bibfnamefont {C.}~\bibnamefont {Kr{\"{u}}ger}},
  \bibinfo {author} {\bibfnamefont {S.}~\bibnamefont {Thutupalli}}, \bibinfo
  {author} {\bibfnamefont {L.}~\bibnamefont {Goehring}}, \ and\ \bibinfo
  {author} {\bibfnamefont {C.}~\bibnamefont {Bahr}},\ }\href {\doibase
  10.1039/c4sm00550c} {\bibfield  {journal} {\bibinfo  {journal} {Soft Matter}\
  }\textbf {\bibinfo {volume} {10}},\ \bibinfo {pages} {7008} (\bibinfo {year}
  {2014})}\BibitemShut {NoStop}%
\bibitem [{\citenamefont {Maass}\ \emph {et~al.}(2016)\citenamefont {Maass},
  \citenamefont {Kr{\"{u}}ger}, \citenamefont {Herminghaus},\ and\
  \citenamefont {Bahr}}]{Maass2016a}%
  \BibitemOpen
  \bibfield  {author} {\bibinfo {author} {\bibfnamefont {C.~C.}\ \bibnamefont
  {Maass}}, \bibinfo {author} {\bibfnamefont {C.}~\bibnamefont {Kr{\"{u}}ger}},
  \bibinfo {author} {\bibfnamefont {S.}~\bibnamefont {Herminghaus}}, \ and\
  \bibinfo {author} {\bibfnamefont {C.}~\bibnamefont {Bahr}},\ }\href {\doibase
  10.1146/annurev-conmatphys-031115-011517} {\bibfield  {journal} {\bibinfo
  {journal} {Annual Review of Condensed Matter Physics}\ }\textbf {\bibinfo
  {volume} {7}},\ \bibinfo {pages} {171} (\bibinfo {year} {2016})}\BibitemShut
  {NoStop}%
\bibitem [{\citenamefont {Illien}\ \emph {et~al.}(2020)\citenamefont {Illien},
  \citenamefont {{De Blois}}, \citenamefont {Liu}, \citenamefont {{Van Der
  Linden}},\ and\ \citenamefont {Dauchot}}]{Illien2020}%
  \BibitemOpen
  \bibfield  {author} {\bibinfo {author} {\bibfnamefont {P.}~\bibnamefont
  {Illien}}, \bibinfo {author} {\bibfnamefont {C.}~\bibnamefont {{De Blois}}},
  \bibinfo {author} {\bibfnamefont {Y.}~\bibnamefont {Liu}}, \bibinfo {author}
  {\bibfnamefont {M.~N.}\ \bibnamefont {{Van Der Linden}}}, \ and\ \bibinfo
  {author} {\bibfnamefont {O.}~\bibnamefont {Dauchot}},\ }\href {\doibase
  10.1103/PhysRevE.101.040602} {\bibfield  {journal} {\bibinfo  {journal}
  {Physical Review E}\ }\textbf {\bibinfo {volume} {101}},\ \bibinfo {pages} {0
  40602} (\bibinfo {year} {2020})}\BibitemShut {NoStop}%
\bibitem [{\citenamefont {{De Buyl}}\ \emph {et~al.}(2013)\citenamefont {{De
  Buyl}}, \citenamefont {Mikhailov},\ and\ \citenamefont
  {Kapral}}]{DeBuyl2013a}%
  \BibitemOpen
  \bibfield  {author} {\bibinfo {author} {\bibfnamefont {P.}~\bibnamefont {{De
  Buyl}}}, \bibinfo {author} {\bibfnamefont {A.~S.}\ \bibnamefont {Mikhailov}},
  \ and\ \bibinfo {author} {\bibfnamefont {R.}~\bibnamefont {Kapral}},\ }\href
  {\doibase 10.1209/0295-5075/103/60009} {\bibfield  {journal} {\bibinfo
  {journal} {Europhys. Lett.}\ }\textbf {\bibinfo {volume} {103}},\ \bibinfo
  {pages} {60009} (\bibinfo {year} {2013})}\BibitemShut {NoStop}%
\bibitem [{\citenamefont {Hyman}\ \emph {et~al.}(2014)\citenamefont {Hyman},
  \citenamefont {Weber},\ and\ \citenamefont {J{\"{u}}licher}}]{Hyman2014}%
  \BibitemOpen
  \bibfield  {author} {\bibinfo {author} {\bibfnamefont {A.~A.}\ \bibnamefont
  {Hyman}}, \bibinfo {author} {\bibfnamefont {C.~A.}\ \bibnamefont {Weber}}, \
  and\ \bibinfo {author} {\bibfnamefont {F.}~\bibnamefont {J{\"{u}}licher}},\
  }\href {\doibase 10.1146/annurev-cellbio-100913-013325} {\bibfield  {journal}
  {\bibinfo  {journal} {Annual Review of Cell and Developmental Biology}\
  }\textbf {\bibinfo {volume} {30}},\ \bibinfo {pages} {39} (\bibinfo {year}
  {2014})}\BibitemShut {NoStop}%
\bibitem [{\citenamefont {Volpe}\ \emph {et~al.}(2011)\citenamefont {Volpe},
  \citenamefont {Buttinoni}, \citenamefont {Vogt}, \citenamefont
  {K{\"{u}}mmerer}, \citenamefont {Bechinger}, \citenamefont {Kuemmerer},\ and\
  \citenamefont {Bechinger}}]{Volpe2011}%
  \BibitemOpen
  \bibfield  {author} {\bibinfo {author} {\bibfnamefont {G.}~\bibnamefont
  {Volpe}}, \bibinfo {author} {\bibfnamefont {I.}~\bibnamefont {Buttinoni}},
  \bibinfo {author} {\bibfnamefont {D.}~\bibnamefont {Vogt}}, \bibinfo {author}
  {\bibfnamefont {H.-J.}\ \bibnamefont {K{\"{u}}mmerer}}, \bibinfo {author}
  {\bibfnamefont {C.}~\bibnamefont {Bechinger}}, \bibinfo {author}
  {\bibfnamefont {H.-J.}\ \bibnamefont {Kuemmerer}}, \ and\ \bibinfo {author}
  {\bibfnamefont {C.}~\bibnamefont {Bechinger}},\ }\href {\doibase
  10.1039/c1sm05960b} {\bibfield  {journal} {\bibinfo  {journal} {Soft Matter}\
  }\textbf {\bibinfo {volume} {7}},\ \bibinfo {pages} {8810} (\bibinfo {year}
  {2011})},\ \Eprint {http://arxiv.org/abs/1104.3203} {arXiv:1104.3203}
  \BibitemShut {NoStop}%
\bibitem [{\citenamefont {Buttinoni}\ \emph {et~al.}(2013)\citenamefont
  {Buttinoni}, \citenamefont {Bialk{\'{e}}}, \citenamefont {K{\"{u}}mmel},
  \citenamefont {L{\"{o}}wen}, \citenamefont {Bechinger},\ and\ \citenamefont
  {Speck}}]{Buttinoni2013}%
  \BibitemOpen
  \bibfield  {author} {\bibinfo {author} {\bibfnamefont {I.}~\bibnamefont
  {Buttinoni}}, \bibinfo {author} {\bibfnamefont {J.}~\bibnamefont
  {Bialk{\'{e}}}}, \bibinfo {author} {\bibfnamefont {F.}~\bibnamefont
  {K{\"{u}}mmel}}, \bibinfo {author} {\bibfnamefont {H.}~\bibnamefont
  {L{\"{o}}wen}}, \bibinfo {author} {\bibfnamefont {C.}~\bibnamefont
  {Bechinger}}, \ and\ \bibinfo {author} {\bibfnamefont {T.}~\bibnamefont
  {Speck}},\ }\href {\doibase 10.1103/PhysRevLett.110.238301} {\bibfield
  {journal} {\bibinfo  {journal} {Physical Review Letters}\ }\textbf {\bibinfo
  {volume} {110}},\ \bibinfo {pages} {238301} (\bibinfo {year}
  {2013})}\BibitemShut {NoStop}%
\bibitem [{\citenamefont {Weeks}\ \emph {et~al.}(1971)\citenamefont {Weeks},
  \citenamefont {Chandler},\ and\ \citenamefont {Andersen}}]{Weeks1971}%
  \BibitemOpen
  \bibfield  {author} {\bibinfo {author} {\bibfnamefont {J.~D.}\ \bibnamefont
  {Weeks}}, \bibinfo {author} {\bibfnamefont {D.}~\bibnamefont {Chandler}}, \
  and\ \bibinfo {author} {\bibfnamefont {H.~C.}\ \bibnamefont {Andersen}},\
  }\href {\doibase 10.1063/1.1674820} {\bibfield  {journal} {\bibinfo
  {journal} {The Journal of Chemical Physics}\ }\textbf {\bibinfo {volume}
  {54}},\ \bibinfo {pages} {5237} (\bibinfo {year} {1971})}\BibitemShut
  {NoStop}%
\bibitem [{\citenamefont {G.~Pellicane}\ and\ \citenamefont
  {Caccamo}(2003)}]{Pellicane}%
  \BibitemOpen
  \bibfield  {author} {\bibinfo {author} {\bibfnamefont {D.~C.}\ \bibnamefont
  {G.~Pellicane}}\ and\ \bibinfo {author} {\bibfnamefont {C.}~\bibnamefont
  {Caccamo}},\ }\href@noop {} {\bibfield  {journal} {\bibinfo  {journal} {J.
  Phys.: Condens. Matter}\ }\textbf {\bibinfo {volume} {15}},\ \bibinfo {pages}
  {375} (\bibinfo {year} {2003})}\BibitemShut {NoStop}%
\bibitem [{\citenamefont {Frenkel}\ and\ \citenamefont {Smit}(2002)}]{Frenkel}%
  \BibitemOpen
  \bibfield  {author} {\bibinfo {author} {\bibfnamefont {D.}~\bibnamefont
  {Frenkel}}\ and\ \bibinfo {author} {\bibfnamefont {B.}~\bibnamefont {Smit}},\
  }\href@noop {} {\emph {\bibinfo {title} {{Understanding Molecular Simulation:
  from algorithms to simulations}}}},\ \bibinfo {edition} {2nd}\ ed.\ (\bibinfo
   {publisher} {Academic Press},\ \bibinfo {address} {Boston},\ \bibinfo {year}
  {2002})\BibitemShut {NoStop}%
\bibitem [{Note1()}]{Note1}%
  \BibitemOpen
  \bibinfo {note} {{If $r_\protect \text {cut}\to \infty $, all the solute
  particles are of type B, and a phase transition occurs in the whole bulk, and
  not only in the reaction area. The fluctuations in the density of B particles
  are not confined to a small region of space anymore, and are not localized at
  the close vicinity of the colloid, therefore preventing the propulsion
  mechanism from occurring.}}\BibitemShut {Stop}%
\bibitem [{\citenamefont {Lippera}\ \emph {et~al.}(2020)\citenamefont
  {Lippera}, \citenamefont {Morozov}, \citenamefont {Benzaquen},\ and\
  \citenamefont {Michelin}}]{Lippera2020b}%
  \BibitemOpen
  \bibfield  {author} {\bibinfo {author} {\bibfnamefont {K.}~\bibnamefont
  {Lippera}}, \bibinfo {author} {\bibfnamefont {M.}~\bibnamefont {Morozov}},
  \bibinfo {author} {\bibfnamefont {M.}~\bibnamefont {Benzaquen}}, \ and\
  \bibinfo {author} {\bibfnamefont {S.}~\bibnamefont {Michelin}},\ }\href
  {\doibase 10.1017/jfm.2019.1055} {\bibfield  {journal} {\bibinfo  {journal}
  {J. Fluid Mech.}\ }\textbf {\bibinfo {volume} {886}},\ \bibinfo {pages} {A17}
  (\bibinfo {year} {2020})}\BibitemShut {NoStop}%
\bibitem [{\citenamefont {Z{\"{o}}ttl}\ and\ \citenamefont
  {Stark}(2014)}]{Zottl2014}%
  \BibitemOpen
  \bibfield  {author} {\bibinfo {author} {\bibfnamefont {A.}~\bibnamefont
  {Z{\"{o}}ttl}}\ and\ \bibinfo {author} {\bibfnamefont {H.}~\bibnamefont
  {Stark}},\ }\href {\doibase 10.1103/PhysRevLett.112.118101} {\bibfield
  {journal} {\bibinfo  {journal} {Phys. Rev. Lett.}\ }\textbf {\bibinfo
  {volume} {112}},\ \bibinfo {pages} {118101} (\bibinfo {year}
  {2014})}\BibitemShut {NoStop}%
\bibitem [{\citenamefont {Matas-Navarro}\ \emph {et~al.}(2014)\citenamefont
  {Matas-Navarro}, \citenamefont {Golestanian}, \citenamefont {Liverpool},\
  and\ \citenamefont {Fielding}}]{Matas-Navarro2014}%
  \BibitemOpen
  \bibfield  {author} {\bibinfo {author} {\bibfnamefont {R.}~\bibnamefont
  {Matas-Navarro}}, \bibinfo {author} {\bibfnamefont {R.}~\bibnamefont
  {Golestanian}}, \bibinfo {author} {\bibfnamefont {T.~B.}\ \bibnamefont
  {Liverpool}}, \ and\ \bibinfo {author} {\bibfnamefont {S.~M.}\ \bibnamefont
  {Fielding}},\ }\href {\doibase 10.1103/PhysRevE.90.032304} {\bibfield
  {journal} {\bibinfo  {journal} {Phys. Rev. E}\ }\textbf {\bibinfo {volume}
  {90}},\ \bibinfo {pages} {032304} (\bibinfo {year} {2014})}\BibitemShut
  {NoStop}%
\bibitem [{\citenamefont {Theers}\ \emph {et~al.}(2018)\citenamefont {Theers},
  \citenamefont {Westphal}, \citenamefont {Qi}, \citenamefont {Winkler},\ and\
  \citenamefont {Gompper}}]{Theers2018}%
  \BibitemOpen
  \bibfield  {author} {\bibinfo {author} {\bibfnamefont {M.}~\bibnamefont
  {Theers}}, \bibinfo {author} {\bibfnamefont {E.}~\bibnamefont {Westphal}},
  \bibinfo {author} {\bibfnamefont {K.}~\bibnamefont {Qi}}, \bibinfo {author}
  {\bibfnamefont {R.~G.}\ \bibnamefont {Winkler}}, \ and\ \bibinfo {author}
  {\bibfnamefont {G.}~\bibnamefont {Gompper}},\ }\href {\doibase
  10.1039/c8sm01390j} {\bibfield  {journal} {\bibinfo  {journal} {Soft Matter}\
  }\textbf {\bibinfo {volume} {14}},\ \bibinfo {pages} {8590} (\bibinfo {year}
  {2018})}\BibitemShut {NoStop}%
\bibitem [{\citenamefont {Weber}\ \emph {et~al.}(2016)\citenamefont {Weber},
  \citenamefont {Weber},\ and\ \citenamefont {Frey}}]{Weber2016}%
  \BibitemOpen
  \bibfield  {author} {\bibinfo {author} {\bibfnamefont {S.~N.}\ \bibnamefont
  {Weber}}, \bibinfo {author} {\bibfnamefont {C.~A.}\ \bibnamefont {Weber}}, \
  and\ \bibinfo {author} {\bibfnamefont {E.}~\bibnamefont {Frey}},\ }\href
  {\doibase 10.1103/PhysRevLett.116.058301} {\bibfield  {journal} {\bibinfo
  {journal} {Phys. Rev. Lett.}\ }\textbf {\bibinfo {volume} {116}},\ \bibinfo
  {pages} {058301} (\bibinfo {year} {2016})}\BibitemShut {NoStop}%
\bibitem [{\citenamefont {Grosberg}\ and\ \citenamefont
  {Joanny}(2015)}]{Grosberg2015}%
  \BibitemOpen
  \bibfield  {author} {\bibinfo {author} {\bibfnamefont {A.~Y.}\ \bibnamefont
  {Grosberg}}\ and\ \bibinfo {author} {\bibfnamefont {J.~F.}\ \bibnamefont
  {Joanny}},\ }\href {\doibase 10.1103/PhysRevE.92.032118} {\bibfield
  {journal} {\bibinfo  {journal} {Phys. Rev. E}\ }\textbf {\bibinfo {volume}
  {92}},\ \bibinfo {pages} {032118} (\bibinfo {year} {2015})}\BibitemShut
  {NoStop}%
\bibitem [{\citenamefont {Ilker}\ and\ \citenamefont
  {Joanny}(2020)}]{Ilker2020}%
  \BibitemOpen
  \bibfield  {author} {\bibinfo {author} {\bibfnamefont {E.}~\bibnamefont
  {Ilker}}\ and\ \bibinfo {author} {\bibfnamefont {J.-F.}\ \bibnamefont
  {Joanny}},\ }\href {\doibase 10.1103/physrevresearch.2.023200} {\bibfield
  {journal} {\bibinfo  {journal} {Physical Review Research}\ }\textbf {\bibinfo
  {volume} {2}},\ \bibinfo {pages} {23200} (\bibinfo {year}
  {2020})}\BibitemShut {NoStop}%
\bibitem [{\citenamefont {Ilker}\ \emph {et~al.}(2021)\citenamefont {Ilker},
  \citenamefont {Castellana},\ and\ \citenamefont {Joanny}}]{Ilker2021}%
  \BibitemOpen
  \bibfield  {author} {\bibinfo {author} {\bibfnamefont {E.}~\bibnamefont
  {Ilker}}, \bibinfo {author} {\bibfnamefont {M.}~\bibnamefont {Castellana}}, \
  and\ \bibinfo {author} {\bibfnamefont {J.-F.}\ \bibnamefont {Joanny}},\
  }\href {http://arxiv.org/abs/2103.06659} {\  (\bibinfo {year} {2021})},\
  \Eprint {http://arxiv.org/abs/2103.06659} {arXiv:2103.06659} \BibitemShut
  {NoStop}%
\end{thebibliography}

%

\end{document}